\documentclass[aps,preprint]{revtex4}%
\usepackage{amsfonts}
\usepackage{amsmath}
\usepackage{amssymb}
\usepackage{graphicx}%
\begin{document}
\title{\textbf{Magnetothermoelectric transport in modulated and unmodulated graphene}}
\author{R. Nasir and K. Sabeeh$^{\dag}$}
\affiliation{Department of Physics,Quaid-i-Azam University, Islamabad $45320$ Pakistan.}
\pacs{PACS number}

\begin{abstract}
We draw motivation from recent experimental studies and present a
comprehensive study of magnetothermoelectric transport in a graphene monolayer
within the linear response regime. We employ the modified Kubo formalism
developed for thermal transport in a magnetic field. Thermopower as well as
thermal conductivity as a function of the gate voltage of a graphene monolayer
in the presence of a magnetic field perpendicular to the graphene plane is
determined for low magnetic fields ($\symbol{126}1$ Tesla) as well as high
fields ($\symbol{126}8$ Tesla). We include the effects of screened charged
impurities on thermal transport. We find good, qualitative as well as
quantitative, agreement with recent experimental work on the subject. In
addition, in order to analyze the effects of modulation, which can be induced
by various means, on the thermal transport in graphene, we evaluate the
thermal transport coefficients for a graphene monolayer subjected to a
periodic electric modulation in a magnetic field. The results are presented as
a function of the magnetic field and the gate voltage.

\end{abstract}
\volumeyear{year}
\volumenumber{number}
\issuenumber{number}
\eid{identifier}
\date[Date text]{date}
\received[Received text]{date}

\revised[Revised text]{date}

\accepted[Accepted text]{date}

\published[Published text]{date}

\startpage{1}
\endpage{2}
\maketitle

\section{\textbf{INTRODUCTION }}

Graphene exhibits remarkable thermal properties. The measured values of
thermal conductivity of graphene reach as high as several thousand of watt per
meter Kelvin\cite{13, 14, 15, 16}, and these are among the highest values of
known materials. Heat transport measures the energy carried by both electrons
and phonons and is fundamental to understanding a material, its ground states,
excitations and scattering mechanisms. If the dream of carbon-based
electronics is to be realized, it is essential to study how and how fast heat
is dissipated across graphene devices. This requires systematic measurements
of thermal conductivity and thermopower over a broad temperature range
(1.5-300 Kelvin) under various external conditions. Therefore, recently there
has been considerable interest, both experimental\cite{1, 2, 3, 4} and
theoretical\cite{5, 6, 7, 8, 9, 10, 11, 12}, in the study of thermoelectric
and magnetothermoelectric transport in graphene. This is partly due to the
realization that the information provided by thermoelectric transport is
complementary to electrical transport. And thermoelectric and
magnetothermoelectric transport studies are extremely useful in providing
insight on the scattering mechanism involved in transport. Fundamentally
related to the electrical conductivity, the thermal conductivity and
thermoelectric coefficients can be determined by the band structure and
scattering mechanisms. The thermoelectric coefficients involve the energy
derivatives of the electrical transport counterparts such as the conductivity
$\sigma$\cite{1}. Recent measurements of thermoelectric power(TEP) on graphene
samples in zero and non-zero magnetic fields have shown a linear temperature
dependence of TEP which suggest that the dominant contribution is that of
diffusive thermopower ($S_{d}$). A comparison between the measured TEP and
that predicted by the Mott formula shows general agreement, particularly at
lower temperatures ($T<50K$)\cite{6}. However, at higher temperatures
deviation from the Mott relation have been reported\cite{2, 3}. In theoretical
work, Yan et al.\cite{11} have determined the TEP of Dirac fermions in
graphene with in the self-consistent Born approximation. Also, Hwang et
al.\cite{5}, in their calculation of TEP incorporate the energy dependence of
various transport scattering rates and show that the dominant contribution is
from the screened charged impurities in graphene's environment. Further,
Vaidya et al.\cite{6} used Boltzmann transport theory to calculate $S_{d}$ in
graphene after considering contributions of optical phonon and surface
roughness scatterings.

Application of a magnetic field in addition to a thermal gradient has profound
effects on the thermal transport in a system and serves as an additional
probe. When a magnetic field is applied perpendicular to the $x-y$ plane of
the sample, the diffusing charge carriers experience the Lorentz force. This
results in developing a transverse electric field $E_{y}$ in addition to the
longitudinal field $E_{x}$. The thermopower is determined from the thermal
gradient $\nabla T$ and the induced voltage $\nabla V$ as $S_{xx}%
=-\frac{\nabla V_{x}}{\nabla T}($ also known as the Seebeck coefficient) and
$S_{xy}=-\frac{\nabla V_{y}}{\nabla T}$(the Nernst coefficient). They are a
measure of the magnitude of the longitudinal and transverse voltages generated
in response to an applied temperature gradient. They are very sensitive in
graphene due to its semimetal nature\cite{8}. The quantum magnetic
oscillations in electrical and thermal transport have been earlier
investigated theoretically by Gusynin and Sharapov\cite{new10} and they
obtained analytical results for longitudinal thermal conductivity and the
Nernst coefficient. However, they assumed a scattering rate that is constant
in energy, independent of magnetic field and temperature. Hence the self
energy used is not self consistent. Moreover, they evaluated the longitudinal
thermal conductivity as a function of the magnetic field at different
temperatures but at fixed chemical potential and constant impurity broadening.
Further, they determined the Nernst coefficient (signal) without recourse to
the modified Kubo formalism appropriate for thermal transport in a magnetic
field. They neglected the dependence of $\Gamma$ on the chemical potential/
carrier concentration. Dora and Thalmeier extended the work presented in
\cite{new10} and studied the electric and thermal response of two dimensional
Dirac fermions in a quantizing magnetic field in the the presence of localized
disorder\cite{new11}.They evaluated the Seeback coefficient and the
corresponding thermal conductivity as a function of the chemical potential and
the magnetic field. They did not determine the Nernst coefficient and the
transverse thermal conductivity.

What distinguishes our work on unmodulated graphene from the aforementioned
previous papers is that we employ the modified Kubo formalism required to
study thermal transport in a magnetic field. As has been discussed earlier,
the usual Kubo formula for thermal response functions is invalid in a magnetic
field and needs to be modified when calculating the transverse (Hall) thermal
conductivity and the Nernst coeffecient\cite{new12,18}. We use the
phenomenological transport equations obtained from the modified Kubo
formalism\cite{18,19}. Further, in the scattering rate and the impurity
broadening of the Landau levels the effects of the carrier concentration that
can be varied by the gate voltage are taken into account. In the first stage,
we determine the components of magnetoelectrothermal(MET) power and MET
conductivity of an unmodulated graphene monolayer in the presence of randomly
distributed charged impurities. The results are presented as a function of the
gate voltage for small and large magnetic fields applied perpendicular to the
graphene sheet. We determine both the Nernst and Seeback coefficients as well
as longitudinal and transverse thermal conductivity. These results are then
compared with experimental work. In addition, we have also carried out a
detailed investigation of the MET transport properties of a graphene monolayer
which is modulated by a weak one-dimensional periodic potential in the
presence of a perpendicular magnetic field. Motivation for this has arisen
from recent work, experimental and theoretical, that has shown that
interaction with a substrate can lead to weak periodic modulation of the
graphene spectrum. Furthermore, applying patterned gate voltage or placing
graphene on a pre-patterned substrate can also lead to modulated
graphene\cite{new7, new8, new9}. Placing impurities or adatom deposition can
do the same. In a previous work, we have computed the electric transport
coefficients of electrically modulated graphene\cite{17}. It was shown that
modulation turns the sharp Landau levels into bands whose width oscillates
periodically with the magnetic field. This affects the magnetoelectric
transport coefficients which exhibit commensurabilty (Weiss) oscillations. The
origin of these Weiss oscillations is the commensurability of the two
characteristic length scales of the system: The cyclotron diameter at the
Fermi energy and the period of the modulation\cite{new1}. An interesting
feature of electronic conduction in the modulated system is the opening of the
diffusive (band) transport channel in addition to hopping (collisional)
transport. Both these contributions to MET transport are taken into account in
this work.

In the following section, we present the general formulation of the
magnetoelectrothermal transport problem and perform the calculation of the
thermopower and the thermal conductivity of unmodulated graphene as well as
graphene subjected to one-dimensional (1D) weak periodic modulation. The
results for the transport coefficients as a function of gate voltage ($V_{g}$)
for unmodulated graphene are discussed in Section III, where we also make a
comparison with experimental results. Following this in Section IV, the
results for modulated graphene as a function of the gate voltage and the
external magnetic field are presented. The present paper ends with a summary
and conclusions.

\section{THERMAL MAGNETOTRANSPORT COEFFICIENTS}

As mentioned in the introduction, corrections to the usual Kubo formula for
transport have to be made when studying thermal transport in a magnetic field.
This was carried out by Luttinger, Smerka, Streda and Oji\cite{19,18}. We
employ the modified Kubo formalism to determine the thermal transport
coefficients from the electrical $J_{e}$ and thermal (energy) current
densities $J_{Q}$
\begin{align}
J_{e\mu}  &  =%
\mathcal{L}%
_{\mu\nu}^{(0)}\left[  -\frac{1}{e}\left(  \nabla_{\nu}\overline{\eta}\right)
\right]  +\frac{%
\mathcal{L}%
_{\mu\nu}^{(1)}}{e}\left[  T\nabla_{\nu}\left(  \frac{1}{T}\right)  \right]
\label{1a}\\
J_{Q\mu}  &  =\frac{%
\mathcal{L}%
_{\mu\nu}^{(1)}}{e}\left[  -\frac{1}{e}\left(  \nabla_{\nu}\overline{\eta
}\right)  \right]  +\frac{%
\mathcal{L}%
_{\mu\nu}^{(2)}}{e^{2}}\left[  T\nabla_{\nu}\left(  \frac{1}{T}\right)
\right]  . \label{1b}%
\end{align}
Here $\overline{\eta}=\eta-e\phi$ with $\eta$ the chemical potential, $\phi$
the scalar potential, $e$ the electronic charge and $T$ the temperature of the
system. The electrical and thermal transport coefficients: the electrical
conductivity $\sigma$, thermopower $S$ and the thermal conductivity $\kappa$
can be obtained from the above expressions, following \cite{18,19, 20, 21,22}, as%

\begin{equation}
\sigma_{\mu\nu}=%
\mathcal{L}%
_{\mu\nu}^{(0)}, \label{1}%
\end{equation}%
\begin{equation}
S_{\mu\nu}=\frac{1}{eT}[(%
\mathcal{L}%
^{(0)})^{-1}%
\mathcal{L}%
^{(1)}]_{\mu\nu}, \label{2}%
\end{equation}%
\begin{equation}
\kappa_{\mu\nu}=\frac{1}{e^{2}T}[%
\mathcal{L}%
_{\mu\nu}^{(2)}-eT(%
\mathcal{L}%
^{(1)}S)_{\mu\nu}] \label{3}%
\end{equation}
with%
\begin{equation}%
\mathcal{L}%
_{\mu\nu}^{(\alpha)}=\int dE\left[  -\frac{\partial f(E)}{\partial E}\right]
(E-\eta)^{\alpha}\sigma_{\mu\nu}(E). \label{4}%
\end{equation}
$%
\mathcal{L}%
_{\mu\nu}^{(\alpha)}$($\alpha=0,1,2$) are, in general, tensors where $\mu
,\nu=x,y$. These phenomenological transport coefficients satisfy the Onsager
relation \cite{19, 20} $%
\mathcal{L}%
_{\mu\nu}^{(\alpha)}(B)=%
\mathcal{L}%
_{\nu\mu}^{(\alpha)}(-B)$. $\sigma_{\mu\nu}(E)$ is the zero-temperature
conductivity and $f(E)=[\exp(\frac{E-\eta}{k_{B}T}+1)]^{-1}$ is the Fermi
Dirac distribution function with $\eta$ the chemical potential. The quantity
$\rho_{\mu\nu}=(%
\mathcal{L}%
^{(0)})_{\mu\nu}^{-1}$ is the resistivity tensor whose components are
$\rho_{xx}=\sigma_{yy}/\Lambda$, $\rho_{yy}=\sigma_{xx}/\Lambda$, $\rho
_{xy}=-\rho_{yx}=\sigma_{yx}/\Lambda$ with $\Lambda=\sigma_{xx}\sigma
_{yy}-\sigma_{xy}\sigma_{yx}$.

In order to calculate the thermal transport coefficients for graphene, we
consider a graphene monolayer in the $xy-plane$ subjected to a magnetic field
$B$ along the $z$-direction. In the Landau gauge, the unperturbed single
particle Dirac-like Hamiltonian may be written as%
\begin{equation}
H_{o}=v_{F}\mathbf{\sigma}.\left(  -i\hbar\mathbf{\nabla}+e\mathbf{A}\right)
. \label{5}%
\end{equation}
Here, $\mathbf{\sigma}=\left\{  \sigma_{x},\sigma_{y}\right\}  $ are the Pauli
matrices and $v_{F}=10^{6}m/s$ characterizes the electron velocity with
$\mathbf{A}=(0,Bx,0)$ the vector potential. The normalized eigenfunctions of
the Hamiltonian given in Eq.(\ref{5}) are%
\begin{equation}
\Psi_{n,k_{y}}=\frac{e^{ik_{y}y}}{\sqrt{2L_{y}l}}\binom{-i\phi_{n-1}\left[
(x+x_{o})/l\right]  }{\phi_{n}\left[  (x+x_{o})/l\right]  }, \label{6}%
\end{equation}
where $\phi_{n}(x)$ and $\phi_{n-1}(x)$ are the harmonic oscillator
wavefunctions centred at $x_{o}=l^{2}k_{y}$. $n$ is the Landau level index,
$l=\sqrt{\frac{\hslash}{eB}}$ the magnetic length and $L_{y}$ the length of 2D
graphene system in the $y$-direction. The corresponding eigenvalue is
$E_{n}=\hslash\omega_{g}\sqrt{n}$ where $\omega_{g}=v_{F}\sqrt{2eB/\hbar
}=v_{F}\sqrt{2}/l$ is the cyclotron frequency of the Dirac electrons in graphene.

In order to investigate the effects of modulation, we express the Hamiltonian
in the presence of modulation as $H=H_{o}+U(x)$. Here, $U(x)$ is the
one-dimensional periodic modulation potential along the $x$-axis. It is given
by $U(x)=V_{e}\cos Kx\ \ $such that $K=\frac{2\pi}{a}$, $a$ is the period of
modulation \ and $V_{e}$ is the constant modulation amplitude. To account for
weak modulation, we take $V_{e}$ to be an order of magnitude smaller than the
Fermi energy $E_{F}=v_{F}\hslash k_{F},$where $k_{F}=\sqrt{2\pi n_{e}}$ is the
magnitude of Fermi wave vector with $n_{e}$ the density of electrons. This
allows us to apply standard first order perturbation theory to determine the
energy eigenvalues in the presence of modulation. Thus, energy eigenvalues for
weak modulation ($V_{e}\ll$ $E_{F}$), are $E_{n,k_{y}}=E_{n}+F_{n,B}\cos Kx$.
Here, $F_{n,B}=\frac{V_{e}}{2}\exp(-\frac{u}{2})[L_{n}(u)+L_{n-1}(u)]$,
$u=\frac{K^{2}l^{2}}{2}$ and, $L_{n}(u)$ and $L_{n-1}(u)\ $are Laguerre polynomials.

In the presence of a periodic modulation, there are two contributions to
magnetoconductivity: the collisional (hopping) contribution and the diffusive
(band) contribution. The former is the localized state contribution which
carries the effects of \ Shubnikov de Hass (SdH) oscillations that are
modified by periodic modulation. The diffusive contribution is the extended
state contribution and arises due to finite drift velocity acquired by the
charge carriers in the presence of modulation. In the linear response regime,
the conductivity tensor is a sum of a diagonal and a non diagonal part :
$\sigma_{\mu\nu}(\omega)=\sigma_{\mu\nu}^{d}(\omega)+\sigma_{\mu\nu}%
^{nd}(\omega)$, $\mu,\nu=x,y$. In general, $\sigma_{\mu\nu}^{d}(\omega
)=\sigma_{\mu\nu}^{diff}(\omega)+\sigma_{\mu\nu}^{\operatorname{col}}%
(\omega),$ accounting for both diffusive and collisional contribution whereas
$\sigma_{\mu\nu}^{nd}(\omega)$ is the Hall contribution. Here, $\sigma
_{xx}=\sigma_{xx}^{\operatorname{col}}$ and $\sigma_{yy}=\sigma_{xx}%
^{\operatorname{col}}+\sigma_{yy}^{diff}.$ Similar to the conductivity
tensors, the diagonal components of the thermal transport coefficients are
determined by the following expressions:%
\begin{equation}%
\mathcal{L}%
_{xx}^{(\alpha)}=%
\mathcal{L}%
_{xx}^{(\alpha)\operatorname{col}}=%
\mathcal{L}%
_{yy}^{(\alpha)\operatorname{col}}\label{10}%
\end{equation}%
\begin{equation}%
\mathcal{L}%
_{yy}^{(\alpha)}=%
\mathcal{L}%
_{yy}^{(\alpha)diff}+%
\mathcal{L}%
_{yy}^{(\alpha)\operatorname{col}}.\label{11}%
\end{equation}
The finite temperature conductivity components $\sigma_{\mu\nu}$ have been
evaluated in \cite{17} for scattering by random screened charged impurities of
density $N_{I}$ with impurity broadening $\Gamma$. The screened potential (in
Fourier space) is $U_{o}=2\pi e^{2}/\epsilon\sqrt{q^{2}+k_{s}^{2},}$ which is
valid for small wave vectors, $q\ll k_{s}$, $k_{s}$ being the inverse
screening length and $\epsilon$ the dielectric constant. Therefore, from
Eq.(\ref{4}), we obtain the zero-temperature phenomenological transport
coefficients $%
\mathcal{L}%
_{\mu\nu}^{(\alpha)}$ as%
\begin{equation}%
\mathcal{L}%
_{yy}^{(\alpha)diff}=2\frac{e^{2}}{h}\frac{\tau}{\hslash}u\underset
{n=0}{\overset{\infty}{\sum}}[F_{n,B}]^{2}[E-\eta]^{\alpha}[\frac{-\partial
f(E)}{\partial E}]_{_{E=E_{n}}},\label{7}%
\end{equation}%
\begin{equation}%
\mathcal{L}%
_{xx}^{(\alpha)\operatorname{col}}\approx\frac{e^{2}}{h}\frac{\beta
N_{I}U_{\circ}^{2}}{\pi a\Gamma}\underset{n=0}{\overset{\infty}{\sum}%
}n\overset{a/l^{2}}{\underset{0}{\int}}dk_{y}[E-\eta]^{\alpha}f_{n,k_{y}%
}(1-f_{n,k_{y}}),\label{8}%
\end{equation}
and%
\begin{equation}%
\mathcal{L}%
_{yx}^{(\alpha)}=\frac{e^{2}}{h}\frac{l^{2}}{a}\underset{n=0}{\overset{\infty
}{\sum}}\overset{a/l^{2}}{\underset{0}{\int}}dk_{y}\frac{1}{\left[  \left(
E_{n+1,k_{y}}-E_{n,k_{y}}\right)  /\hslash\omega_{g}\right]  ^{2}}%
\overset{E_{n+1,k_{y}}}{\underset{E_{n,k_{y}}}{\int}}dE\left\{  [E-\eta
]^{\alpha}[\frac{-\partial f(E)}{\partial E}]\right\}  _{E_{n,k_{y}}%
},\label{9}%
\end{equation}
where $\tau$ is the scattering time. Here, we have taken the scattering time
to be independent of Landau-level index $n$. And the components of thermopower
are given by the following equations:%
\begin{equation}
S_{xx}=\frac{1}{eT}\left[  \left(  \frac{\sigma_{yy}}{S_{o}}\right)
\mathcal{L}%
_{xx}^{(1)}+\left(  \frac{1}{\sigma_{yx}}\right)
\mathcal{L}%
_{yx}^{(1)}\right]  ,\label{12}%
\end{equation}%
\begin{equation}
S_{yy}=\frac{1}{eT}\left[  \left(  \frac{\sigma_{xx}}{S_{o}}\right)
\mathcal{L}%
_{yy}^{(1)}+\left(  \frac{1}{\sigma_{yx}}\right)
\mathcal{L}%
_{yx}^{(1)}\right]  \label{13}%
\end{equation}
and%
\begin{equation}
S_{xy}=\frac{1}{eT}\left[  \left(  \frac{\sigma_{yy}}{S_{o}}\right)  (-%
\mathcal{L}%
_{yx}^{(1)})+\left(  \frac{1}{\sigma_{yx}}\right)
\mathcal{L}%
_{yy}^{(1)}\right]  ,\label{14}%
\end{equation}%
\begin{equation}
S_{yx}=\frac{1}{eT}\left[  \left(  \frac{\sigma_{xx}}{S_{o}}\right)
\mathcal{L}%
_{yx}^{(1)}+\left(  -\frac{1}{\sigma_{yx}}\right)
\mathcal{L}%
_{xx}^{(1)}\right]  .\label{15}%
\end{equation}
The components of the thermal conductivity are given by%
\begin{equation}
\kappa_{xx}=\frac{1}{e^{2}T}\left[
\mathcal{L}%
_{xx}^{(2)}-eT\left\{
\mathcal{L}%
_{xx}^{(1)}S_{xx}-%
\mathcal{L}%
_{yx}^{(1)}S_{yx}\right\}  \right]  ,\label{16}%
\end{equation}%
\begin{equation}
\kappa_{yy}=\frac{1}{e^{2}T}\left[
\mathcal{L}%
_{yy}^{(2)}-eT\left\{
\mathcal{L}%
_{yx}^{(1)}S_{xy}+%
\mathcal{L}%
_{yy}^{(1)}S_{yy}\right\}  \right]  \label{17}%
\end{equation}
and%
\begin{equation}
\kappa_{xy}=\frac{1}{e^{2}T}\left[  -%
\mathcal{L}%
_{yx}^{(2)}-eT\left\{
\mathcal{L}%
_{xx}^{(1)}S_{xy}-%
\mathcal{L}%
_{yx}^{(1)}S_{yy}\right\}  \right]  ,\label{18}%
\end{equation}%
\begin{equation}
\kappa_{yx}=\frac{1}{e^{2}T}\left[
\mathcal{L}%
_{yx}^{(2)}-eT\left\{
\mathcal{L}%
_{yx}^{(1)}S_{xx}+%
\mathcal{L}%
_{yy}^{(1)}S_{yx}\right\}  \right]  .\label{19}%
\end{equation}

\section{MAGNETOTHERMOPOWER AND MAGNETOTHERMAL CONDUCTIVITY OF
UNMODULATED\ GRAPHENE}

From the electrical conductivity $\sigma_{\mu\nu},$ calculated in our previous
work \cite{17}, we determine the phenomenological transport coefficients $%
\mathcal{L}%
_{yx}^{(\alpha)}$ employing Eqs.\ref{7}, \ref{8} and \ref{9}. Employing these,
the components of thermopower and thermal conductivity are numerically
evaluated using Eqs.\ref{12} through \ref{19}. The results for the
magnetoelectrothermal transport properties of an unmodulated graphene
monolayer as a function of the gate voltage are presented in this section. The
number density $n_{e\text{ }}$is related to the gate voltage $V_{g}$ through
the relationship $n_{e}=\epsilon_{o}\epsilon V_{g}/te$, where $\epsilon_{o}$
and $\epsilon=3.9$ are the permittivities for free space and the dielectric
constant for graphene on a SiO$_{2}$ substrate, respectively. The electron
charge is $e$ and $t(\approx300nm)$ is the thickness of the sample \cite{23}.
The components of thermopower ($S_{\mu\nu}$) and thermal conductivity
($\kappa_{\mu\nu}$), as the system moves away from the charge neutral point on
the electron side on changing the gate voltage, are shown in Fig. (1) at a
magnetic field of one Tesla. The lattice temperature of $10K$ \ and mobility
of $\mu=20m^{2}/Vs$\cite{new3} is chosen. The scattering time is related to
the mobility as $\tau=\frac{\mu E_{F}}{ev_{F}^{2}}$ in a graphene
monolayer\cite{new4}. Since impurity broadening $\Gamma$ can be expressed in
terms of the self energy $\Sigma^{-}(E)$ as $\Gamma\equiv\Gamma
(E)=2\operatorname{Im}\left[  \Sigma^{-}(E)\right]  $ and also $\Gamma
(E)=\hslash/\tau$\cite{new5}. We use the expression for $\operatorname{Im}%
\left[  \Sigma^{-}(E)\right]  $ derived in \cite{17} to find $\Gamma
=\sqrt{\hslash(\hslash\omega_{g})^{2}/(4\pi\tau E_{F})}$. The electron number
density is $n_{e\text{ }}=7.19V_{g}\times10^{14}m^{-2}$ and Fermi energy is
$E_{F}=\hbar v_{F}\sqrt{2\pi n_{e}}=44.3\sqrt{V_{g}}meV$. And the impurity
density is related to $\Gamma$ through $N_{I}=\pi l^{2}\Gamma^{2}/U_{o}^{2}%
$\cite{new6}. The scattering time of $\tau=4.431\mu\sqrt{V_{g}}\times
10^{-14}s$, impurity broadening $\Gamma=5.934\sqrt{B/(\mu V_{g})}meV$\ and
impurity density $N_{I}=\frac{2.46}{\mu}\times10^{14}m^{-2}$ were employed in
this work\cite{new3, new4, new5, new6, 24, 25, 26}. Moreover, the same study
is carried out at a higher magnetic field of $8.8T$ for graphene with
mobilities of $\mu=1m^{2}/Vs$ and $\mu=20m^{2}/Vs$ respectively and the
results are shown in Fig. (2). Since $S_{xx}$ and $S_{yy}$ are identical so
only $S_{xx}$ is depicted in these figures. The longitudinal coefficient of
thermopower ($S_{xx}$) is equivalent to the Seebeck coefficient and our
results provide a qualitative as well as quantitative understanding of the
overall behavior of the observed $S_{xx}(V_{g})$. $S_{xx}$ can have either
sign and it is negative in our case since the charge carriers are electrons in
this range of $V_{g}$.\ The transverse component of thermopower ($S_{yx}$) is
also known as the Nernst signal and it arises due to the presence of the
perpendicular magnetic field as the Lorentz force bends the trajectories of
the thermally diffusing carriers. It can be seen from Fig. (1a) and Fig. (2a)
that $S_{xx}$ follows $1/\sqrt{V_{g}}$ (with $V_{g}\propto n_{e}$). Similar
behavior of $S_{xx}$ is observed in experiments\cite{1, 2, 3}. Notice that we
have presented results for diffusive thermopower and we have ignored the
phonon contribution to thermopower due to weak electron phonon coupling in
graphene \cite{2, 5}. $S_{xx}$ and $S_{yx}$ ( $S_{yx}=-S_{xy}$ )\ both show
Shubnikov-de Haas (SdH) type oscillations in the Landau quantizing magnetic
field. At the lower magnetic field of 1 Tesla (Fig. (1)), the oscillations are
more closely spaced since the separation between the Landau levels, which is
proportional to the magnetic field strength, is smaller compared to the
results for the higher magnetic field of 8.8 Tesla, (Fig. (2)). Moreover, we
observe in Figs. (1a) and (2a) that both $S_{xx}$ and $S_{yx}$ approach zero
at those values of $V_{g}$ where there are boundaries of Landau Levels and no
carriers are available to participate in transport. The peaks of $S_{xx}$ are
observed at the centre of Landau levels. With the increase in $V_{g}$ and
hence an increase in $n_{e}$, higher Landau levels are occupied. The
oscillations in $S_{xx}$ and $S_{xy}$ are damped as we increase $V_{g}$. The
reason for this is that higher $V_{g}$ corresponds to higher values of the
Fermi energy and if the Fermi energy is much larger that the Landau level
separation, Landau quantization effects are lost. At $B=8.8T,$ (Fig. (2a)),
the oscillations in $S_{xx}$ and $S_{xy}$ show that the width of the peaks
broaden compared to those for smaller magnetic field of $B=1T$. At lower
magnetic field, the separation between the Landau levels is smaller compared
to higher fields with the result that the peaks of $S_{xx}$ are more closely
spaced. Furthermore, the overall magnitude of $S_{xx}$ and \ $S_{xy}$
increases with increasing magnetic field strength (See Fig. 1a and Fig.
2a).\ In these figures, we also present thermal conductivity as a function of
the gate voltage. The longitudinal thermal conductivity $\kappa_{xx}$ shows
oscillating behavior which damps out as $V_{g}$ increases, where Landau
quantization effects become less significant. However, the transverse
component of thermal conductivity $\kappa_{yx}$ rises monotonically with
$V_{g}$ as shown in Fig. (1b) and Fig. (2b). At the higher magnetic field,
quantum Hall steps have begun to appear. The behavior of longitudinal and
transverse thermal conductivity follows that of the corresponding components
of electrical conductivity. At higher magnetic fields, the splitting of the
peaks in the longitudinal thermal condutivity $\kappa_{xx}$ is seen in Fig.
(2b) which was also observed in \cite{new11} where it is shown that the
splitting occurs in such a way that they produce antiphase oscillations with
respect to the electric one and lead to the violation of the Wiedemann-Franz
law. For the un-modulated case, $\sigma_{yy}=\sigma_{xx}$ , $%
\mathcal{L}%
_{xx}^{(\alpha)}=%
\mathcal{L}%
_{yy}^{(\alpha)}$ and using Eq.(\ref{12}) through Eq.(\ref{19}) we find that
$S_{yx}=-S_{xy}$, $\kappa_{xx}=\kappa_{yy}$ and $\kappa_{xy}=-\kappa_{yx}$.
Therefore, only $\kappa_{xx}$ and $\kappa_{yx}$ are shown in the figures. We
find that the results for magnetothermal power obtained in our work at
$B=8.8T$ with $T=10K$ are in good agreement, both qualitative and
quantitative, with the experimental results obtained in \cite{2, 3}, see Fig
(3) of \cite{2}. These results do indicate that scattering from screened
charged impurities is the dominant scattering mechanism required to explain
the experimental results. We must add that our quantitative results for
$S_{yx}$ depend strongly on the mobility of the graphene system.

\section{MAGNETOTHERMOPOWER AND MAGNETOTHERMAL CONDUCTIVITY OF PERIODICALLY
MODULATED GRAPHENE}

Now we consider the effects of modulation. The 1D modulation broadens the
sharp Landau levels into bands and gives rise to an additional diffusive (or
band) contribution to transport. This additional contribution is absent
without modulation. We now focus on the modulation induced changes in the
thermal magnetotransport coefficients of graphene. Therefore, in the first
part we present the thermopower and thermal conductivity of modulated graphene
with mobility of $20m^{2}/Vs$ as a function of the gate voltage. These are
shown in Fig. (3) and Fig. (5) respectively. The results are for a constant
external magnetic field of $B=1T$ applied perpendicular to the graphene sheet,
with electric modulation of strength $V_{e}=3meV$ applied in the $x$-direction
at a temperature of $T=10K$. In this case $\Gamma=\frac{1.3}{\sqrt{V_{g}}}meV$
and $\hslash\omega_{g}=36.3meV$, such that $\Gamma\ll V_{e}\ll\hslash
\omega_{g}$ to satisfy the requirements of weak modulation. The period of
modulation is $a=382nm$. The results for $S_{xx}$ and $S_{yy}$ are identical,
so only $S_{xx}$ is shown in these figures. The amplitude of oscillations in
$S_{xx}$ ($\Delta S_{xx}$) is greater than that of $S_{xy}$ ($\Delta S_{xy}$)
which damps out with increasing gate voltage($V_{g}$). Both $S_{xx}$ and
$S_{xy}$ show SdH-type oscillations and it verifies that the system is Landau
quantized. The modulations effects are apparent in $S_{xx}$ and $S_{xy}$ which
shows modulation of SdH-type oscillations and $\Delta S_{xx}\gg\Delta S_{xy}$,
Fig. (3). $\kappa_{yx}$ is greater than $\kappa_{xx}$ and $\kappa_{yy}$ as
shown in Fig. (5). These modulation induced effects on thermal transport
coefficients can be highlighted by calculating the difference between the
modulated case and the un-modulated case. The contribution of modulation to
thermopower $\Delta S_{\mu\nu}(V_{e})=\Delta S_{\mu\nu}(V_{e})-\Delta
S_{\mu\nu}(0)$ and thermal conductivity $\Delta\kappa_{\mu\nu}(V_{g}%
)=\Delta\kappa_{\mu\nu}(V_{e})-\Delta\kappa_{\mu\nu}(0)$ are shown in Fig. (4)
and Fig. (6) respectively. These figures clearly show the modulation of SdH
oscillations in both the thermopower and the thermal conductivity. For an
un-modulated case $\kappa_{xx}=\kappa_{yy}$, however for modulated graphene
$\kappa_{xx}\neq\kappa_{yy}$ and this expected behaviour is seen in Fig. (6)
where $\Delta\kappa_{xx}\neq\Delta\kappa_{yy}$. The 1D modulation gives a
positive contribution to $\Delta\kappa_{yy}$ while $\Delta\kappa_{xx}$ and
$\Delta\kappa_{yx}$ oscillate around zero. $\Delta\kappa_{yy}\gg\Delta
\kappa_{xx}$, which is a consequence of the fact that $\Delta\kappa_{xx}$ has
only collisional contribution, whereas $\Delta\kappa_{yy}$ in addition to the
collisional part, has large contribution from band conduction.

We also show the the results when the magnetic field is varied and the
electron density is fixed at $n_{e}=3.16\times10^{15}m^{-2}$ which corresponds
to a gate voltage of $V_{g}=4.39V$. The Fermi energy of the system is
$E_{F}=\hbar v_{F}\sqrt{2\pi n_{e}}\approx92.3meV$. We have taken the mobility
of $20m^{2}/Vs$\cite{new3}and hence scattering time is taken to be
$\tau=1.86\times10^{-12}$ $s$. Impurity broadening $\Gamma=0.633\sqrt{B}meV$
and impurity density $N_{I}=1.23\times10^{13}m^{-2}$ were employed in this
part of the work. The strength of the electrical modulation is taken to be
$V_{e}=2meV$ with period $a=382nm$ and temperature $T=10K$. The difference
between the modulated case and the unmodulated case highlights the modulation
induced effects in these thermoelectric quantities. The thermopower and the
change in thermopower due to modulation $\Delta S_{\mu\nu}(B)$ are shown in
Fig. (7) as a function of the magnetic field in the units of $-k_{B}/e$. When
$B$ is less than $0.2T$ Weiss oscillations are observed whereas SdH type
oscillations dominate at higher magnetic fields. It is also seen that these
oscillations in $S_{xx}$ are $90^{o}$ out of phase with those in $S_{xy}.$ The
amplitude of the oscillations in $\Delta S_{xx}\gg\Delta S_{xy}$ and they are
$90^{o}$ out of phase. Again for $B$ greater than $0.2T$ the oscillations
appear as envelopes of SdH oscillations. The different components of the
thermal conductivity tensor and the correction to it due to 1D modulation are
shown in Fig. (8). The magnetic field dependence of the thermal conductivity
tensor is similar to that of the electrical conductivity tensor obtained in
\cite{17}. In Fig. (8) we see that $\Delta\kappa_{yx}\gg\Delta\kappa_{yy}%
\gg\Delta\kappa_{xx},$ such that $\Delta\kappa_{yx}$ and $\Delta\kappa_{xx}$
are $180^{o}$ out of phase from each other.

To conclude, in this work we have studied magnetothermoelectric transport in
graphene in the linear response regime using the modified Kubo formalism
appropriate for thermal transport in a magnetic field. Results are presented
for both unmodulated graphene as well as graphene that is weakly modulated by
an electric modulation. We take into account scattering from screened charged
impurities and our results indicate that these provide the most dominant
scattering mechanism at low temperatures. The thermopower, the Seebeck
coefficient and the Nernst coefficient are determined as a function of the
gate voltage. Furthermore, we also determine the magnetothermal conductivity
tensor, both the longitudinal and the transverse (Hall) components. For
unmodulated graphene we were able to make a comparison of the thermopower with
experimental results and find that they are in good agreement, both
qualitative as well as quantitative, with experimental results. In the case of
modulated graphene, we focus on the modulation induced effects that appear as
commensurabilty (Weiss-type) oscillations in the magnetothermoelectric
coefficients. The results are presented as both functions of the gate voltage
and the magnetic field.

\section{Acknowledgements}

R. Nasir and K. Sabeeh would like to acknowledge the support of the Higher
Education Commission (HEC) of Pakistan through project No. 20-1484/R\&D/09.

$\dagger$ Corresponding author: ksabeeh@qau.edu.pk \

\end{document}